\documentclass[12pt]{article}

\usepackage[english]{babel}
\usepackage{latexsym}
\usepackage{epsfig}

\begin{document}

\title{Delocalisation phenomena in one-dimensional models with
       long-range correlated disorder: a perturbative approach}

\author{L.~Tessieri \\
{\it Department of Chemistry, Simon Fraser University,} \\
{\it Burnaby, British Columbia, Canada V5A 1S6}}

\date{18th June 2002}

\maketitle

\begin{abstract}
We study the nature of electronic states in one-dimensional
continuous models with weak correlated disorder. Using a perturbative
approach, we compute the inverse localisation length (Lyapunov
exponent) up to terms proportional to the fourth power of the
potential; this makes possible to analyse the delocalisation
transition which takes place when the disorder exhibits specific
long-range correlations. We find that the transition consists
in a change of the Lyapunov exponent, which switches from a quadratic
to a quartic dependence on the strength of the disorder.
Within the framework of the fourth-order approximation, we also
discuss the different localisation properties which distinguish
Gaussian from non-Gaussian random potentials.
\end{abstract}

{Pacs numbers:  73.20.Jc, 72.15.Rn, 72.10.Bg}

\section{Introduction}

In recent years, the interest for one-dimensional (1D) models with
correlated disorder has been steadily growing, as it has become
progressively clear that correlations of the random potential can
deeply affect the structure of the electronic eigenstates and
endow 1D disordered models with far richer transport properties than
it was previously thought.

For a long time it was believed that 1D systems could not display
complex features like the metal-insulator transition which takes place
in three-dimensional (3D) models, since it was known that all
one-electron states are localised in 1D systems with totally random
potentials, regardless of the strength of disorder (see~\cite{Lif88}
and references therein).
Further research, however, led to the study of some variants of the
standard 1D Anderson model which exhibited a {\em discrete} set of
extended states (see, e.g.,~\cite{Flo89}). All these systems were
characterised by random potentials with {\em short-range} spatial
correlations, a feature absent from the original Anderson model where
the site energies are totally uncorrelated.

Eventually, the role of {\em long-range} correlations was also
investigated and it was shown that this kind of correlations can
produce a {\em continuum} of extended states~\cite{Mou98}.
This finding was corroborated by the work of Izrailev and Krokhin who,
using second-order perturbation theory, managed to establish an
analytical relation between localisation length and potential pair
correlators, and used this result to show how specific long-range
disorder correlations lead to the appearance of {\em mobility edges}
in 1D discrete models~\cite{Izr99}.
An experimental confirmation of this result was obtained by studying
the transmission of microwaves in a single-mode waveguide with a
random array of correlated scatterers~\cite{Kuh00}.
The results established in~\cite{Mou98} and~\cite{Izr99} for
discrete lattices were subsequently extended to continuous models
and related to parallel phenomena occurring in different fields like
the propagation of waves in random media~\cite{Izr01} and the dynamics
of classical stochastic oscillators~\cite{Tes01}.

The discovery that 1D disordered systems can display a metal-insulator
transition analogous to the one which takes place in 3D models
constituted a crucial advancement for the understanding of anomalous
transport properties of 1D random models. This theoretical progress
is also relevant from the technological point of view, because it
paves the way for the construction of 1D devices with pre-determined
mobility edges which can be used as window filters in electronic,
acoustic, and photonic structures~\cite{Iku01}.
The importance of the result makes therefore highly desirable to
reach a complete comprehension of the link between long-range
correlations of the disorder and the appearance of a continuum of
extended electronic states in 1D models.

At the analytical level, our understanding of this problem rests on
the results which were first derived in~\cite{Izr99}.
Unfortunately, both this work and the ones that have followed in its
wake suffered from two main limitations: the disorder was supposed to
be weak, so that a perturbative approach could be applied, and all the
analytical results were obtained in the second-order approximation.
(These constraints were absent from the work~\cite{Mou98} of de Moura
and Lyra, but they drew their conclusions from numerical calculations
and did not provide any analytical insight on the delocalisation
mechanism.)
Analytical calculations are greatly simplified by truncating the
expansion of the inverse localisation length to the second-order term;
the main drawback of this choice is that it leaves open the question
of the true nature of the states of the ``extended'' phase.
In fact, a vanishing second-order inverse localisation length can be
related to various physical phenomena: it can indicate that the
electronic states are completely delocalised, but it can also be a sign
of weak forms of localisation, characterised by power-law decay of the
electronic probability distribution; finally, it is the result to be
expected when the spatial ranges of electronic states in the ``extended''
and ``localised'' phases differ by orders of magnitude.

The present work constitutes an attempt to partially remove the limits
of the previous analyses; in this study we still focus our attention on
the case of {\em weak} disorder but, with the help of a systematic
perturbative technique, we manage to go beyond the second order
approximation and obtain analytical results correct to the fourth
order of perturbation theory.
By this way we can shed light on the true nature of the states that are
classified as ``extended'' in the framework of the second-order theory
and we are also able to discuss the differences which emerge at this
refined level of description between Gaussian and non-Gaussian random
potentials.

In the case of Gaussian disorder with long-range correlations, it turns
out that the electronic states are exponentially localised on both sides
of the mobility edge identified in~\cite{Izr99}; however, the dependence
of their inverse localisation length on the disorder strength changes
from a {\em quadratic} to a {\em quartic} form upon crossing the
critical threshold.
For weak disorder, this implies that electrons in the ``extended'' states
can actually move over far longer distances than the electrons which
find themselves in a ``localised'' state.
The use of terms like ``mobility edge'' and ``delocalisation transition''
is therefore legitimate, provided that one keeps in mind that the 
qualification of ``extended'' must be understood as ``localised on a
large spatial scale'' rather than ``completely delocalised''.
The analysis of the localisation length in the fourth-order approximation
also reveals that, upon increasing the energy of the electrons, a second
threshold appears, which separates the states whose spatial extension
scales with the inverse of the fourth power of the potential from
states that are characterised by an even weaker localisation (whose exact
form cannot be ascertained within the limits of the fourth-order
approximation).

The distinction between Gaussian and non-Gaussian disorder cannot
be discussed within the framework of the second-order approximation,
because the second-order inverse localisation length only depends on
the second moment of the random potential, and differences between
Gaussian and non-Gaussian distributions only show up in the higher
moments.
A systematic computation of the higher-order terms of the inverse
localisation length, however, establishes a connection between the
$n$-th term of the expansion and the corresponding $n$-th moment of
the potential: therefore, the fourth-order results obtained in this
work allow us a first analysis of the differences between Gaussian
and non-Gaussian disorder.
Lifting the Gaussian requirement seems to strengthen the localisation
of electronic states, because the fourth-order inverse localisation
length is increased by a new term proportional to the fourth
{\em cumulant} of the random potential (cumulants are specific
combinations of the moments which vanish in the Gaussian case but are
otherwise non-zero).
Specifically, we analyse in detail the difference between a Gaussian
potential with long-range binary correlations and a non-Gaussian
potential with the same pair correlators but with a fourth-order
cumulant having exponential decreasing form: we find that the non-zero
cumulant produces quartic localisation of {\em all} electronic states
that lie beyond the second-order mobility edge, thus wiping out the
additional threshold that appears in the Gaussian case.

This paper is organised as follows: in Sec.~\ref{probfor} we give an
exact formulation of the problem and we define the model under study; in
Sec.~\ref{pertexp} we expose the perturbative method used and we give
the general results for the second- and fourth-order terms of the
localisation length. We devote Sec.~\ref{src} to the case of random
potential with short-range correlations; the problem of long-range
correlation is then discussed in Sec.~\ref{lrc}. Finally, we summarise
the main results and express our conclusions in Sec.~\ref{conclu}.

\section{Formulation of the problem}
\label{probfor}

\subsection{Definition of the model}

We consider the 1D disordered model defined by the stationary
Schr\"{o}dinger equation
\begin{equation}
- \frac{\hbar^{2}}{2m} \psi''(x) + \varepsilon U(x) \psi(x) =
E \psi(x)
\label{mod}
\end{equation}
where $\psi(x)$ represents the state of a quantum particle
(``electron'') of mass $m$ and energy $E$ moving in a continuous
random potential $U(x)$. The dimensionless parameter $\varepsilon$
is introduced to measure the strength of the disorder; for the sake
of simplicity, in the rest of this paper we will adopt a system of
units in which $\hbar^{2}/2m =1$. We will set the zero of the energy
scale at the average value of the potential and we will assume that
$E>0$ on this scale.

To complete the definition of the system under study, the statistical
properties of the random potential must be precised. In the first place,
we will assume that model~(\ref{mod}) is spatially homogeneous in the
mean; from the mathematical point of view, this requirement implies
that the moments of $U(x)$ satisfy the invariance condition
\begin{equation}
\langle U(x_{1} +\delta) U(x_{2} +\delta) \ldots U(x_{n} +\delta)
\rangle = \langle U(x_{1}) U(x_{2}) \ldots U(x_{n}) \rangle
\label{traninv}
\end{equation}
for every spatial translation $\delta$. In Eq.~(\ref{traninv}), and
throughout this paper, the angular brackets denote the average over
disorder realisations, or ensemble average.
Property~(\ref{traninv}) implies that the $n$-th moment of the random
potential can depend only on the relative positions of the points
$(x_{1},x_{2},\ldots,x_{n})$; this allows one to represent the
$n$-point correlator as a function of $n-1$ relative coordinates
according to the identity
\begin{equation}
\chi_{n} (\tilde{x}_1,\tilde{x}_2, \ldots, \tilde{x}_{n-1}) =
\langle U(x) U(x+\tilde{x}_1) U(x+\tilde{x}_2) \ldots
U(x+\tilde{x}_{n-1}) \rangle .
\label{corfun}
\end{equation}
Our second basic assumption on model~(\ref{mod}) is that its average
features do not depend on the sign of the random potential, i.e.,
that the physical properties of the model are invariant under the
transformation $U(x) \rightarrow -U(x)$. This is ensured by assuming
that all odd moments of the potential vanish
\begin{displaymath}
\langle U(x_1) U(x_2) \ldots U(x_{2n+1}) \rangle = 0 .
\end{displaymath}
Finally, for reasons of mathematical convenience, we will suppose that
a finite length scale $l_{c}$ exists for the model~(\ref{mod}) such that
the statistical correlations between the values $U(x_{1})$ and $U(x_{2})$
of the potential become negligible when the distance separating the
points $x_{1}$ and $x_{2}$ exceeds $l_{c}$. For the particular case of
the two-point correlation function this assumption translates in the
condition
\begin{equation}
\begin{array}{ccc}
\langle U(x_{1}) U(x_{2}) \rangle \simeq 0 & \mbox{for} &
| x_{1} - x_{2} | \gg l_{c} \\
\end{array} ;
\label{lc}
\end{equation}
more generally, the existence of a correlation length $l_{c}$ implies
that the n-point correlation functions $\langle U(x_{1}) U(x_{2}) \ldots
U(x_{n})$ must satisfy the so-called ``product property'', meaning that,
if the points $x_{1},x_{2}, \ldots, x_{n}$ can be divided in two groups
with $|x_{i} - x_{j} | \gg l_{c}$ for $x_{i}$ and $x_{j}$ belonging to
different groups, the average $\langle U(x_{1}) U(x_{2}) \ldots U(x_{n})$
takes the value obtained by averaging the two different groups separately.
To deal with the case of long-range correlations, we will first make use
of the previous hypothesis and derive results valid for any finite $l_{c}$;
when possible, we will then take the limit $l_{c} \rightarrow \infty$.

Having thus defined the statistical properties of the random potential,
we can proceed to enunciate the most specific assumption about the
model under investigation. In the following, we will restrict our
considerations to the case of {\em weak} disorder, as defined by the
relation
\begin{equation}
\varepsilon^{2} \langle U(x) U(x') \rangle \ll E^{2} .
\label{weakdis}
\end{equation}
and by the analogous conditions on the higher moments of the random
potential.
Notice that, to ensure that condition~(\ref{weakdis}) is valid
independently of the specific form of the function $U(x)$, the parameter
$\varepsilon$ must satisfy the condition $\varepsilon \ll 1$.

Eq.~(\ref{mod}), together with the specified statistical properties of
the random potential and the assumption of weak disorder, completely
defines the model under study. Our goal consists in determining
the spatial behaviour of the solutions of Eq.~(\ref{mod}); to this end
it is necessary to give a precise definition of the key parameter
known as localisation length.

\subsection{Localisation length}
\label{lldefin}

To shed light on the localisation properties of model~(\ref{mod}), we
will use as an indicator of the localised or extended nature of the
electronic states the quantity
\begin{equation}
\lambda = \lim_{x \rightarrow \infty} \frac{1}{4x}
\ln \langle \psi^{2}(x) k^{2} + \psi'^{2}(x) \rangle ,
\label{loclen}
\end{equation}
where the parameter $k$ is related to the electronic energy by
the relation $E = k^{2}$.
In definition~(\ref{loclen}) the wavefunction $\psi(x)$ can be supposed
to be real, so that no need arises to consider the complex extension
of the logarithm function.

In this paper we will always refer to the quantity~(\ref{loclen}) as the
{\em inverse localisation length} or {\em Lyapunov exponent}, although
this parameter is more usually defined through the expression
\begin{equation}
\tilde{\lambda} = \lim_{x \rightarrow \infty} \frac{1}{2x}
\langle \ln \left( \psi^{2}(x) k^{2} + \psi'^{2}(x) \right) \rangle .
\label{stdlen}
\end{equation}
The parameters~(\ref{loclen}) and~(\ref{stdlen}) differ in principle,
but both can be used effectively to ascertain the localised or extended
character of the electronic states: in fact, they belong to the same
family of generalised Lyapunov exponents
\begin{displaymath}
L_{q} = \lim_{x \rightarrow \infty} \frac{1}{x}
\ln \left( \langle |\psi(x)|^{q} \rangle^{1/q} \right) ,
\end{displaymath}
which were introduced long ago for the study of localisation in 1D
disordered models (see~\cite{Pal87} and references therein).
In technical terms, Eq.~(\ref{stdlen}) defines the standard Lyapunov
exponent $L_{0}$, whereas the parameter~(\ref{loclen}) is equal to the
``generalised Lyapunov exponent of order two'' divided by a factor two,
$\lambda = L_{2}/2$.
Our preference for the definition~(\ref{loclen}) of the inverse
localisation length over the possible alternative~(\ref{stdlen})
stems mainly from the fact that the first choice makes the study of
model~(\ref{mod}) more amenable to analytical treatment.
The exact relation between the Lyapunov exponents~(\ref{loclen})
and~(\ref{stdlen}) is not a completely solved problem; however, it is
clear that the two affine indicators can be used equivalently to
determine whether an eigenfunction of Eq.~(\ref{mod}) is localised or
not. Indeed, for delocalised states the alternative
definitions~(\ref{loclen}) and~(\ref{stdlen}) give the same result,
since both parameters vanish when the solutions of Eq.~(\ref{mod})
are extended.
As for exponentially localised states, for the case of interest here,
i.e., that of weak disorder, one can prove that the Lyapunov
exponents~(\ref{loclen}) and~(\ref{stdlen}) coincide at least to the
second order of perturbation theory~\cite{Tes01}. In Sec.~\ref{src}
we will come back to this point to show that the identity seems to
hold also beyond the second-order approximation.

From the physical point of view, the difference between the
alternative parameters~(\ref{loclen}) and~(\ref{stdlen}) is perhaps
best understood if the problem of electronic localisation in
model~(\ref{mod}) is re-defined in terms of the energetic instability
of a stochastic oscillator.
In fact, the physical comprehension of the model~(\ref{mod}) can be
enhanced by observing that solving the stationary Schr\"{o}dinger
equation~(\ref{mod}) is completely equivalent to studying the dynamics
of the classical stochastic oscillator defined by the Hamiltonian
\begin{equation}
H = k \left( \frac{p^{2}}{2} + \frac{q^{2}}{2} \right) +
\varepsilon k \frac{q^{2}}{2} \xi(t) .
\label{randosc}
\end{equation}
The Hamiltonian~(\ref{randosc}) represents an oscillator with
frequency $k$ perturbed by a noise $\xi(t)$; the equivalence of
this system with model~(\ref{mod}) can be seen by considering that the
dynamical equation of the oscillator
\begin{equation}
\ddot{q}(t) + k^{2} \left( 1 + \varepsilon \xi(t) \right)
q(t) = 0
\label{dyneq}
\end{equation}
coincides with Eq.~(\ref{mod}) if one identifies the electronic 
wavefunction $\psi(x)$ with the orbit $q(t)$ of the oscillator and if
the noise $\xi(t)$ is related to the random potential $U(x)$ by the
identity
\begin{equation}
\xi(x) = - U(x)/E.
\label{nois}
\end{equation}

Underlying the mathematical equivalence of Eqs.~(\ref{mod})
and~(\ref{dyneq}) is the physical parallelism between the localisation
of the quantum states of model~(\ref{mod}) and the energetic
instability of the stochastic oscillator~(\ref{randosc})~\cite{Tes01,
Tes00}.
The connection between the two phenomena emerges clearly if one writes
the inverse localisation length~(\ref{loclen}) in terms of the dynamical
variables of the random oscillator. The Lyapunov exponent~(\ref{loclen})
can then be expressed as
\begin{displaymath}
\lambda = \lim_{t \rightarrow \infty} \frac{1}{4t}
\ln \langle k^{2} \left( p^{2}(t) + q^{2}(t) \right) \rangle
\end{displaymath}
and interpreted as the growth rate of the average energy of the
random oscillator~(\ref{randosc}). A positive value of the
parameter~(\ref{loclen}) can therefore be read both as a sign of
exponential localisation of the eigenstates of model~(\ref{mod}) and
as an indication that the stochastic oscillator~(\ref{randosc}) is
energetically unstable.

As for parameter~(\ref{stdlen}), it can be rewritten in terms of
the oscillator variables as
\begin{displaymath}
\tilde{\lambda} = \lim_{T \rightarrow \infty}
\lim_{\delta \rightarrow 0}
\int_{0}^{T} \ln \frac{x(t+\delta)}{x(t)} dt .
\end{displaymath}
In this form, the parameter~(\ref{stdlen}) reveals itself as the
exponential divergence rate of nearby orbits, i.e., as the Lyapunov
exponent for the trajectories of the oscillator~(\ref{randosc}).
We can conclude that the parameters~(\ref{loclen}) and~(\ref{stdlen})
are both growth rates, measuring respectively the increase of the
oscillator energy and that of the distance between initially nearby
orbits.

\section{Perturbative expansion and general results}
\label{pertexp}

To compute the localisation length~(\ref{loclen}), one needs to
determine the asymptotic behaviour of $\langle \psi^{2}(x) \rangle$
and $\langle \psi'^{2}(x) \rangle$.
A convenient way to achieve this goal consists in deriving a
deterministic differential equation for the average squares of the
wavefunction and of its derivative. To this end, we introduce the
scaled derivative
\begin{displaymath}
\phi(x) = \psi'(x)/k
\end{displaymath}
and the vector
\begin{displaymath}
u(x) = \left( \begin{array}{c} \psi^{2}(x) \\
                               \phi^{2} (x) \\
                               \psi(x) \phi(x)
              \end{array} \right) .
\end{displaymath}
Taking the Schr\"{o}dinger equation~(\ref{mod}) as starting point, it
is easy to prove that $u(x)$ obeys the stochastic equation
\begin{equation}
\frac{d u}{dx} =
\left( {\bf{A}} + \varepsilon \xi(x) {\bf{B}} \right) u ,
\label{stoceq}
\end{equation}
where $\xi(x)$ is the scaled potential~(\ref{nois}) and the symbols
$\bf{A}$ and $\bf{B}$ stand for the matrices
\begin{displaymath}
\begin{array}{ccc}
\bf{A} = \left( \begin{array}{ccc} 0 &  0 &  2k \\
                                   0 &  0 & -2k \\
                                  -k &  k &  0
                \end{array}
         \right) &
\mbox{and} &
\bf{B} = \left( \begin{array}{ccc} 0 & 0 &  0 \\
                                   0 & 0 & -2k \\
                                  -k & 0 &  0
                \end{array}
         \right) .
\end{array}
\end{displaymath}
We now need to replace the stochastic equation~(\ref{stoceq}) with an
ordinary differential equation for $\langle u(x) \rangle$; this can
be done in the following way. For a fixed initial condition
$u(0) = u_{0}$, the average solution of Eq.~(\ref{stoceq}) can be
formally expressed as
\begin{equation}
\langle u(x) \rangle = {\bf{Z}}(x) u_{0}
\label{zed1}
\end{equation}
with
\begin{displaymath}
{\bf{Z}}(x) = \exp \left( {\bf{A}} x \right) \left[ {\bf{1}} +
\langle
\mbox{T} \! \exp \int_{0}^{x} \varepsilon {\bf M}(x') dx'
\rangle \right] ,
\end{displaymath}
where $\mbox{T} \! \exp$ indicates the T-ordered exponential and
${\bf M}(x)$ is the random matrix
\begin{equation}
{\bf{M}}(x) = \xi(x) \exp \left(-{\bf{A}}x\right) {\bf{B}}
\exp \left( {\bf{A}} x \right) .
\label{ranmat}
\end{equation}
Differentiating both sides of Eq.~(\ref{zed1}) one obtains
\begin{equation}
\frac{d \langle u \rangle}{dx} = {\bf{Z}}'(x) u_{0} ;
\label{zed2}
\end{equation}
on the other hand, by inverting the relation~(\ref{zed1}), one
arrives at
\begin{displaymath}
u_{0} = {\bf{Z}}^{-1}(x) \langle u(x) \rangle .
\end{displaymath}
Substituting this result in the right-hand side (r.h.s.) of
Eq.~(\ref{zed2}), one finally obtains the desired equation for the
average of $u(x)$
\begin{equation}
\frac{d \langle u(x) \rangle}{dx} = {\bf{Z}}'(x)
{\bf{Z}}^{-1}(x) \langle u(x) \rangle
= {\bf{K}}(x) \langle u(x) \rangle .
\label{nonstoc}
\end{equation}

For weak disorder ($\varepsilon \rightarrow 0$), the generator
${\bf{K}}(x)$ that appears in Eq.~(\ref{nonstoc}) can be represented
by an expansion in powers of the perturbative parameter~$\varepsilon$
\begin{equation}
{\bf{K}}(x) = \sum_{n=0}^{\infty} \varepsilon^{n} {\bf{K}}_{n}(x) .
\label{cumexp}
\end{equation}
Van Kampen~\cite{Kam74} and Terwiel~\cite{Ter74} have shown that all
partial generators ${\bf{K}}_{n}(x)$ can be expressed in terms of
specific combinations of the moments of the random matrix~(\ref{ranmat})
known as ``ordered cumulants'' and they have provided a systematic
set of rules to construct them.
Applying Van Kampen's prescriptions to the present case, the zeroth-order
generator turns out to have the simple form
\begin{displaymath}
{\bf{K}}_{0} = {\bf{A}}
\end{displaymath}
which corresponds to neglecting altogether the effects of the weak
random potential.
As for the first-order term, it vanishes like all odd-order generators
${\bf{K}}_{2n+1}(x)$ because of the condition that the odd moments of
$U(x)$ be null.
The second-order generator is less trivial and represents the first
term where the effects of the random potential appear
\begin{equation}
{\bf{K}}_{2} (x) = \exp \left( {\bf{A}} x \right)
\int_{0}^{x} dx_{1} \langle {\bf{M}}(x) {\bf{M}}(x_1) \rangle
\exp \left( -{\bf{A}} x \right) .
\label{k2}
\end{equation}
Straightforward passages allow to derive from this compact expression the
matrix elements of ${\bf{K}}_{2}(x)$, whose explicit form is
\begin{eqnarray}
\left( {\bf{K}}_{2} \right)_{11}(x) & = &
\left( {\bf{K}}_{2} \right)_{12}(x) =
\left( {\bf{K}}_{2} \right)_{13}(x) =
\left( {\bf{K}}_{2} \right)_{23}(x) =
\left( {\bf{K}}_{2} \right)_{32}(x) = 0 \nonumber \\
\left( {\bf{K}}_{2} \right)_{21}(x) & = &
\frac{1}{k^{2}} \int_{0}^{x}
\chi_{2}(y) \left[ 1 + \cos \left( 2ky \right) \right] dy
\nonumber \\
\left( {\bf{K}}_{2} \right)_{22}(x) & = &
\left( {\bf{K}}_{2} \right)_{33}(x) =
\frac{1}{k^{2}} \int_{0}^{x}
\chi_{2}(y) \left[ -1 + \cos \left( 2ky \right) \right] dy
\nonumber \\
\left( {\bf{K}}_{2} \right)_{31}(x) & = & \frac{1}{k^{2}}
\int_{0}^{x} \chi_{2}(y) \sin \left( 2ky \right) dy
\label{k231}
\end{eqnarray}
where the symbol $\chi_{2}$ represents the two-point correlator of the
potential $U(x)$ defined by Eq.~(\ref{corfun}).
The fourth-order generator has a more complex form and depends both on
the four- and the two-point correlators of the potential
\begin{equation}
\begin{array}{l}
{\bf{K}}_{4} (x) =
\int_{0}^{x} dx_1 \int_{0}^{x_1} dx_2 \int_{0}^{x_2} dx_3
\exp \left( -{\bf{A}} x \right) \\
\times \left[
\langle {\bf{M}}(x) {\bf{M}}(x_{1}) {\bf{M}}(x_{2}) {\bf{M}}(x_{3})
\rangle \right. \\
- \langle {\bf{M}}(x) {\bf{M}}(x_{1}) \rangle
\langle {\bf{M}}(x_{2}) {\bf{M}}(x_{3}) \rangle \\
- \langle {\bf{M}}(x) {\bf{M}}(x_{2}) \rangle
\langle {\bf{M}}(x_{1}) {\bf{M}}(x_{3}) \rangle \\
- \left. \langle {\bf{M}}(x) {\bf{M}}(x_{3}) \rangle
\langle {\bf{M}}(x_{1}) {\bf{M}}(x_{2}) \rangle \right]
\exp \left( {\bf{A}} x \right)
\end{array}
\label{k4}
\end{equation}
The integrand in Eq.~(\ref{k4}) is a non-trivial example of {\em ordered
cumulant}; notice that it differs from an ordinary cumulant because the
matrices ${\bf{M}}(x_{1})$ and ${\bf{M}}(x_{2})$ do not commute for
$x_{1} \neq x_{2}$.
The explicit form of the matrix elements of ${\bf{K}}_{4} (x)$ can
be worked out from Eq.~(\ref{k4}); the expressions, however, are quite
lengthy and do not add much to the physical understanding of the problem,
so that we will omit them here.
In theory, the Van Kampen scheme allows one to compute the terms
of the expansion~(\ref{cumexp}) to any desired order; in practice,
however, the involved calculations become rather cumbersome beyond the
second order. In this paper we will not consider terms of higher order
than the fourth; within this approximation, we will be able to
derive the standard second-order localisation length and the first
correction to this result.

In principle, the generators ${\bf{K}}_{n}(x)$ are functions of the
spatial coordinate $x$ (and of the initial condition $u(0)$ chosen to
solve Eq.~(\ref{stoceq})), but the dependence on these factors dies out
as soon as the condition $x \gg l_{c}$ is fulfilled~\cite{Kam74}.
To determine the asymptotic behaviour of $\langle u(x) \rangle$,
therefore, we need not bother with the space-dependent generators
${\bf{K}}_{n}(x)$ but we have only to consider their constant
asymptotic limits
\begin{equation}
{\bf{K}}_{n} = \lim_{x \rightarrow \infty} {\bf{K}}_{n} (x) .
\label{asymgen}
\end{equation}
Notice that the existence of the asymptotic generators~(\ref{asymgen})
is mathematically ensured by the sufficient condition that the random
potential $U(x)$ possess a finite correlation length $l_{c}$.
This implies that, when we relax this condition in order to study the
case of long-range correlations, we will have to pay attention on whether
the asymptotic generators~(\ref{asymgen}) continue to be well-defined in
the limit $l_{c} \rightarrow \infty$.

To sum up, on the basis of the previous considerations we can conclude
that, in order to analyse with fourth-order accuracy the asymptotic
behaviour of $\langle u(x) \rangle$, we can legitimately replace in
Eq.~(\ref{nonstoc}) the exact generator with its asymptotic and truncated
form
\begin{equation}
{\bf{K}}(x) \simeq {\bf{A}} + \varepsilon^{2} {\bf{K}}_{2} +
\varepsilon^{4} {\bf{K}}_{4} .
\label{truncgen}
\end{equation}
The eigenvalue of the matrix~(\ref{truncgen}) with the largest real
part determines the exponential growth rate of the vector $\langle u(x)
\rangle$ and, consequently, the inverse localisation
length~(\ref{loclen}).
The original problem is thus reduced to that of solving the secular
equation of the matrix~(\ref{truncgen}); determining the largest root
of this equation, one obtains that the desired inverse localisation
length has the form
\begin{displaymath}
\lambda = \varepsilon^{2} \lambda_{2} + \varepsilon^{4} \lambda_{4} +
o(\varepsilon^{4})
\end{displaymath}
and that the general expressions for the terms $\lambda_{2}$ and
$\lambda_{4}$ are
\begin{equation}
\lambda_{2}(k) =
\frac{1}{4 k^{2}} \int_{0}^{\infty} \chi_{2} (x) \cos(2kx) dx
\label{lambda2}
\end{equation}
and
\begin{equation}
\begin{array}{ccl}
\displaystyle \lambda_{4}(k) & = &
\frac{1}{8 k^{5}} \int_{0}^{\infty} dx_1 \chi_{2}(x_{1})
\int_{0}^{\infty} dx_2  \chi_{2}(x_{2}) \sin(2kx_2) \\
& + &
\frac{1}{4 k^{4}} \lim_{x \rightarrow \infty} \int_{\omega(x)} dx_{1}
dx_{2} dx_{3}
\left\{ \left[ \chi_{4}(x_{1},x_{1}+x_{2},x_{1}+x_{2}+x_{3})
\right. \right. \\
& - & \left. \chi_{2}(x_{1}) \chi_{2}(x_{3}) \right]
\left[ \cos(2kx_{1}) \cos(2kx_{3}) - \cos \left( 2kx_{1}+2kx_{2}+2kx_{3}
\right) \right] \\
& - &
\chi_{2}(x_{1}+x_{2}) \chi_{2}(x_{2}+x_{3}) \left[ \cos \left(
2kx_{1} + 2kx_{2} \right) \cos \left( 2kx_{2} + 2kx_{3} \right)
\right.\\
& - &
\left. \cos \left( 2kx_{1}+2kx_{2}+2kx_{3} \right) \right] -
\chi_{2}(x_{1}+x_{2}+x_{3}) \chi_{2}(x_{2}) \\
& \times & \left.
\left[ \cos \left( 2kx_{1}+2kx_{2}+2kx_{3} \right) \cos
\left( 2kx_{2} \right) - \cos \left( 2kx_{1}+2kx_{2} \right) \right]
\right\}
\end{array}
\label{lambda4}
\end{equation}
where $\omega(x)$ is the integration domain
\begin{displaymath}
\omega(x) = \left\{ (x_{1},x_{2},x_{3}) : 0 \le x_{1}, 0 \le x_{2},
0 \le  x_{3}, x_{1}+x_{2}+x_{3} \le x \right\} .
\end{displaymath}

Expression~(\ref{lambda2}) reproduces the standard second-order formula
for the Lyapunov exponent of model~(\ref{mod}) (see, e.g.,~\cite{Lif88});
the result shows that the second-order inverse localisation length for
an eigenstate with wavevector $k$ is proportional to the cosine transform
of the binary potential correlator taken at twice the value of the wave
vector.
Not surprisingly, the fourth-order Lyapunov exponent~(\ref{lambda4}) has
a more complicated form, involving both second and fourth moments of the
potential. The increasing complexity of the mathematical formulae reflects
the different degree of sophistication of the second- and fourth-order
approximation schemes. From the physical point of view, this difference
is perhaps better understood if one visualises localisation as an
interference effect (see, e.g.,~\cite{Kam76}).
In the second-order (or Born) approximation, localisation is seen as
an interference phenomenon generated by double scatterings of the
electron at two points separated by a distance over which the random
potential has non-negligible correlations: hence the connection
between the second-order localisation length and the two-point
correlator established by formula~(\ref{lambda2}). The picture
corresponding to the second-order approximation, therefore, includes
an electronic wavefunction (with wavevector $k$) which is scattered first
backward, generating a wave of intensity $O(\varepsilon)$ and then is
rescattered in the forward direction, with intensity $O(\varepsilon^{2})$.
To go beyond the Born approximation, one must include the interference
effects produced by multiple scatterings in the analysis of
localisation. The fourth-order approximation, for instance, takes into
account both quadruple scatterings and couples of double scatterings.
In the first case, an electronic wavefunction is scattered back and forth
four times, generating a forward-scattered wave of intensity
$O(\varepsilon^{4})$ which interferes constructively with the incoming
wave; in the case of double scatterings, one considers the interference
between two twice-scattered waves, each of intensity $O(\varepsilon^{2})$.
This is the reason why the term~(\ref{lambda4}) involves the four-point
potential correlator as well as various combinations of double products
of two-points correlators.

To make the result for the fourth-order term~(\ref{lambda4})
mathematically simpler and physically more transparent, it is
opportune to express the fourth moment of the potential in terms
of the second moments and of the fourth-order cumulant, writing
\begin{equation}
\begin{array}{lcl}
\chi_{4}(x_{1},x_{2},x_{3}) & = & \chi_{2}(x_{1}) \chi_{2}(x_{3}-x_{2})
+ \chi_{2}(x_{2}) \chi_{2}(x_{3}-x_{1}) \\
& + & \chi_{2}(x_{3}) \chi_{2}(x_{2}-x_{1})
+ \Delta_{4}(x_{1},x_{2},x_{3})
\end{array}
\label{chi4}
\end{equation}
where we have used the symbol $\Delta_{4}$ to represent the fourth
cumulant.
Eq.~(\ref{chi4}) represents the generalisation for non-Gaussian
disorder of the familiar identity which allows one to express the
fourth moment in terms of those of order two in the Gaussian case.
In fact, in expression~(\ref{chi4}) the fourth moment is given as
the sum of two parts: the Gaussian combination of the second moments
and the cumulant, which represents the non-Gaussian contribution and
vanishes in the Gaussian case.
This interpretation of Eq.~(\ref{chi4}) makes quite natural to
distinguish two components in the fourth-order Lyapunov exponent and
to write
\begin{equation}
\lambda_{4} (k) = \lambda_{4}^{(G)} (k) + \lambda_{4}^{(NG)} (k)
\label{l4}
\end{equation}
where the first term in the r.h.s. of Eq.~(\ref{l4}) represents the
Gaussian part which depends only on the second moments of the random
potential, whereas the second term is proportional to the cumulant
and constitutes the specific manifestation of the non-Gaussian nature
of disorder on the localisation properties of model~(\ref{mod}).
Inserting~(\ref{chi4}) in formula~(\ref{lambda4}) one obtains that
the explicit forms of the Gaussian and non-Gaussian terms are
\begin{equation}
\begin{array}{l}
\lambda_{4}^{(G)} (k) = \frac{1}{8 k^{5}}
\int_{0}^{\infty} dx_1 \chi_{2}(x_{1}) \int_{0}^{\infty} dx_2
\chi_{2}(x_{2}) \sin(2kx_2) \\
+ \frac{1}{4 k^{4}}
\lim_{x \rightarrow \infty} \int_{\omega (x)} dx_1 dx_2 dx_3
\left\{ \chi_{2} (x_{1}+x_{2}) \chi_{2} (x_{2}+x_{3}) \right. \\
\times
\left[ \cos(2kx_{1}) \cos(2kx_{3}) - \cos \left( 2kx_{1}+2kx_{2}
\right) \cos \left( 2kx_{2}+2kx_{3} \right) \right] \\
+ \chi_{2}(x_{2}) \chi_{2}(x_{1}+x_{2}+x_{3}) \left[ \cos \left( 2kx_{1}
\right) \cos \left( 2kx_{3} \right) \right. \\
+ \cos \left( 2kx_{1} + 2kx_{2} \right)
- \left. \left. \cos \left( 2kx_{1}+2kx_{2}+2kx_{3} \right) \left( 1 +
\cos \left( 2kx_{2} \right) \right) \right] \right\}
\end{array}
\label{gauloc}
\end{equation}
and
\begin{equation}
\begin{array}{ccl}
\lambda_{4}^{(NG)} (k) & = & \displaystyle \frac{1}{4 k^{4}}
\lim_{x \rightarrow \infty} \int_{\omega (x)} dx_1 dx_2 dx_3
\Delta_{4} (x_{1},x_{1}+x_{2},x_{1}+x_{2}+x_{3}) \\
& \times & \left[ \cos (2kx_{1})
\cos (2kx_{3}) - \cos (2kx_{1} + 2kx_{2} + 2kx_{3}) \right] . \\
\end{array}
\label{ngauloc}
\end{equation}
The Gaussian term~(\ref{gauloc}) can be further simplified with
some calculus work, provided that the binary correlation function decays
quickly enough at infinity, which is ensured by condition~(\ref{lc}).
In this way one can lower the dimensionality of the integral~(\ref{gauloc}),
reducing it to a much more manageable one-dimensional integral.
Introducing the functions
\begin{equation}
\begin{array}{ccc}
\varphi_{c}(k,x)  & = &
\displaystyle \int_{x}^{\infty} \chi_{2}(y) \cos(2ky) dy \\
\varphi_{s}(k,x) & = &
\displaystyle \int_{x}^{\infty} \chi_{2}(y) \sin(2ky) dy \\
\end{array},
\label{phifun}
\end{equation}
the final expression for the Gaussian part of the fourth-order
Lyapunov exponent can be written in the form
\begin{equation}
\begin{array}{ccl}
\displaystyle \lambda_{4}^{(G)} (k) & = &
\displaystyle
\frac{1}{16 k^{5}} \varphi_{s}(k,0) \left[ \varphi_{c}(k,0) +
4 \varphi_{c}(0,0) \right] \\
& - & \displaystyle
\frac{1}{8 k^{4}} \left[ \varphi_{c}(k,0) + 2 \varphi_{c}(0,0) \right]
\int_{0}^{\infty} \chi_{2}(x) x \cos (2kx) dx \\
& + & \displaystyle
\frac{1}{8k^{4}} \varphi_{s}(k,0)
\int_{0}^{\infty} \chi_{2}(x) x \sin (2kx) dx \\
& + & \displaystyle
\frac{1}{4k^{4}} \int_{0}^{\infty} \left[ \varphi_{c}(0,x) \varphi_{c}(k,x)
- \varphi_{c}^{2}(0,x) \cos(2kx) \right] dx \\
\end{array}
\label{l4g}
\end{equation}

As for the non-Gaussian term~(\ref{ngauloc}), the dimensionality of
the integral cannot be lowered without formulating some specific
hypothesis on the form of the cumulant $\Delta_{4}$. A change of
variable, however, allows one to cast expression~(\ref{ngauloc})
in the form
\begin{equation}
\begin{array}{ccl}
\lambda_{4}^{(NG)} (k) & = & \displaystyle
\frac{1}{k^{4}} \int_{0}^{\infty} dx
\int_{0}^{x} dy \int_{x}^{\infty} dz \Delta_{4} (y,x,z) \\
& \times & \displaystyle
\left[ \cos(2ky) \cos(2kz-2kx) - \cos(2kz) \right] \\
\end{array}
\label{l4ng}
\end{equation}
which is more convenient for our later use.

Formulae~(\ref{l4g}) and~(\ref{l4ng}) constitute the central result of
this section. They represent the general form of the Gaussian and
non-Gaussian parts of the fourth-order Lyapunov exponent~(\ref{l4});
by applying them to specific cases, we will be able to perform an
accurate analysis of the localisation properties of
model~(\ref{mod}).

\section{Disorder with short range correlations}
\label{src}

In this section, we analyse the localisation properties of random
potentials characterised by two different kinds of short-range
correlations:  a delta-correlated potential (white noise) and 
a potential with exponentially decaying correlations.

\subsection{White noise}
A white noise is defined as a random potential whose cumulants have
the form
\begin{equation}
\Delta_{n} (x_{1},x_{2},\ldots,x_{n-1}) =
\sigma_{n} \delta(x_{1}) \delta(x_{2}) \ldots \delta(x_{n-1})
\label{whicum}
\end{equation}
(see, e.g.,~\cite{Kam92}).
In Eq.~(\ref{whicum}) the $\sigma_{n}$ are constants which determine
the strength of the cumulants, while $(x_{1},x_{2},\ldots,x_{n-1})$
represent relative coordinates of the points
$(x,x+x_{1},x+x_{2} \ldots,x+x_{n-1})$.

Let us consider the Gaussian case first. In this case, all cumulants
vanish except the second, which coincides with the correlation
function
\begin{equation}
\Delta_{2} (x) = \chi_{2}(x) = \sigma_{2} \delta(x) .
\label{whichi}
\end{equation}
Substituting the binary correlator~(\ref{whichi}) in Eq.~(\ref{lambda2}),
one recovers the standard result for the second-order inverse
localisation length
\begin{equation}
\lambda_{2}(k) = \frac{\sigma_{2}}{8 k^{2}} .
\label{whil2}
\end{equation}
To determine the fourth-order correction to this result, the first
step consists in observing that in the case of white noise the
functions~(\ref{phifun}) take the form
\begin{equation}
\begin{array}{ccl}
\varphi_{c} (x) & = & \left\{ \begin{array}{ccc}
                                   0       & \mbox{for} & x>0 \\
                              \sigma_{2}/2 & \mbox{for} & x=0 \\
                              \sigma_{2}   & \mbox{for} & x<0 \\
                              \end{array} \right. \\
\varphi_{s} (x) & = & 0 \\
\end{array}
\label{whiphi}
\end{equation}
Substituting~(\ref{whichi}) and~(\ref{whiphi}) in expression~(\ref{l4g}),
one arrives at the conclusion that the fourth-order Gaussian
correction is zero
\begin{displaymath}
\lambda_{4}^{(G)} (k) = 0 .
\end{displaymath}

This peculiar result is a consequence of the delta-like nature of the
correlation function.
To understand this point, one needs to go back to Eq.~(\ref{k4}) which
shows that the partial generator ${\bf{K}}_{4}(x)$ is an integral
of the fourth ordered cumulant of the random matrix~(\ref{ranmat}).
As pointed out in Sec.~\ref{pertexp}, an ordered cumulant differs from
an ordinary one because the matrices ${\bf{M}}(x)$ do not commute
for different values of $x$. In the special case when the correlation
function has the form~(\ref{whichi}), however, the integrand in
Eq.~(\ref{k4}) vanishes unless all integration variables are equal
to $x$. Consequently, the matrices in~(\ref{k4}) commute and the
ordered cumulant reduces to an ordinary one, which is zero because of
the Gaussian assumption.
Actually, this reasoning can be extended to all generators
${\bf{K}}_{n}(x)$ with $n>2$ and one is thus led to the conclusion that,
in the special case of Gaussian white noise, the asymptotic limit of the
generator~(\ref{cumexp}) is exactly equal to
\begin{equation}
{\bf{K}} = {\bf{A}} + \varepsilon^{2} {\bf{K}}_{2} .
\label{whik}
\end{equation}
The fact that ${\bf{K}}_{4}(x)$ vanishes does not automatically
imply that the fourth-order Lyapunov exponent must be zero; however
it is possible to prove that this is actually the case by solving the
secular equation for the exact generator~(\ref{whik}).
In this way one arrives at the result
\begin{displaymath}
\lambda =
\varepsilon^{2} \frac{\sigma_{2}}{8 k^{2}} -
\varepsilon^{6} \frac{\sigma_{2}^{3}}{128 k^{6}}
+ o(\varepsilon^{6}) 
\end{displaymath}
which shows that no term of order $O(\varepsilon^{4})$ arises from
the second-order generator ${\bf{K}}_{2}$ in the case of Gaussian
white noise.

If we drop the Gaussian assumption, the fourth-order generator no
longer vanishes; however, assuming that the fourth cumulant has the
form~(\ref{whicum}), one finds that the non-Gaussian term~(\ref{l4ng})
is zero.
We are thus led to the conclusion that, in the case of white noise,
all eigenstates of Eq.~(\ref{mod}) are localised, because the
second-order Lyapunov exponent~(\ref{whil2}) is strictly positive for
all values of the wavevector $k$. However, the specific nature of the
correlations prevents quadruple scatterings of electrons from producing
any net destructive or constructive interference effect on the
wavefunction, so that neither an enhancement nor a reduction of
localisation show up in the fourth-order approximation.

\subsection{Exponentially decaying correlations}

We now turn our attention to the case of disorder with exponentially
decaying correlations. Specifically, we assume that the two-point
correlation function and the fourth-order cumulant have the form
\begin{equation}
\chi_{2} (x) = \sigma_{2} \exp \left( - \beta_{2} |x| \right)
\label{expchi}
\end{equation}
and
\begin{equation}
\Delta_{4} (x,y,z) = \sigma_{4} \exp \left[ - \beta_{4} \left( |x|
+ |y| + |z| \right) \right] .
\label{expcum}
\end{equation}
In Eqs.~(\ref{expchi}) and~(\ref{expcum}), the constants $\beta_{2}^{-1}$
and $\beta_{4}^{-1}$ represent the range of the correlation function and
of the fourth cumulant, respectively; note that the two parameters may be
different.

Inserting the binary correlator~(\ref{expchi}) in Eq.~(\ref{lambda2}),
one obtains that the second-order Lyapunov exponent is
\begin{equation}
\lambda_{2}(k) = \frac{\sigma_{2}}{4k^{2}}
\frac{\beta_{2}}{4k^{2}+ \beta_{2}^{2}} .
\label{expl2}
\end{equation}
This result confirms the general rule that random potentials with
short-range correlations produce localisation of all the electronic
eigenstates (with the possible exception of a discrete set of extended
states).
Note that the Lyapunov exponent~(\ref{expl2}) tends to zero in the
limit $\beta_{2} \rightarrow 0$: physically, this means that the
localisation of the electronic states becomes weaker and weaker
as the range of the disorder correlations stretches over
increasingly larger distances.
In the opposite case, i.e. when the range of the correlation function
tends to zero, one can recover from Eq.~(\ref{expl2}) the
result~(\ref{whil2}) already derived for the white-noise case. More
precisely, one obtains the white-noise expression if the limit
$\beta_{2} \rightarrow \infty$ is taken while keeping constant the
integral of the second moment:
$\int_{-\infty}^{+\infty} \chi_{2}(x) dx = \tilde{\sigma_{2}}$.
This condition implies that the correlation strength $\sigma_{2}$ must
scale as $\sigma_{2} = \tilde{\sigma_{2}} \beta_{2}/2$; when this
relation is satisfied, the second-order Lyapunov exponent~(\ref{expl2})
reduces to the white-noise form~(\ref{whil2}) for
$\beta_{2} \rightarrow \infty$.

The binary correlator~(\ref{expchi}) also determines the form of the
Gaussian part of the fourth-order Lyapunov exponent.
For $x > 0$ the functions~(\ref{phifun}) take the form
\begin{eqnarray*}
\varphi_{c} (k,x) & = & \frac{\sigma_{2}}{4k^{2} + \beta_{2}^{2}}
\left[ \beta_{2} \cos (2kx) - 2k \sin (2kx) \right]
\exp \left(- \beta_{2} x \right) \\
\varphi_{s} (k,x) & = & \frac{\sigma_{2}}{4k^{2} + \beta_{2}^{2}}
\left[ 2k \cos (2kx) + \beta_{2} \sin (2kx) \right]
\exp \left(- \beta_{2} x \right) ; \\
\end{eqnarray*}
correspondingly, Eq.~(\ref{l4g}) becomes
\begin{equation}
\lambda_{4}^{(G)}(k) = \frac{\sigma_{2}^{2}}{4}
\frac{\beta_{2} \left( \beta_{2}^{4} + 22k^{2}\beta_{2}^{2} + 48k^{4}
\right)}{k^{4} \left( \beta_{2}^{2} + k^{2} \right) \left( \beta_{2}^{2}
+ 4k^{2} \right)^{3}}
\label{expl4g}
\end{equation}
This result coincides with the expression obtained in Ref.~\cite{Han89}
for the fourth-order term of the inverse localisation
length~(\ref{stdlen}); an important consequence of Eq.~(\ref{expl4g}),
therefore, is that the identity of the localisation lengths~(\ref{loclen})
and~(\ref{stdlen}) is not restricted to the second order, but holds at
least up to the fourth order when disorder is weak and Gaussian and the
correlation function has the exponential form~(\ref{expchi}).
This coincidence further corroborates the conclusions of
subsection~\ref{lldefin} about the substantial equivalence of the
generalised Lyapunov exponents~(\ref{loclen}) and~(\ref{stdlen}).

To complete the discussion of the fourth-order approximation, we have
to compute the non-Gaussian term~(\ref{l4ng}) when the fourth cumulant
has the form~(\ref{expcum}). A straightforward calculation leads to
\begin{equation}
\lambda_{4}^{(NG)}(k) = \frac{15\sigma_{4}}{8} \frac{\beta_{4}}{k^{2}
(\beta_{4}^{2} + k^{2}) (\beta_{4}^{2} + 4k^{2}) (9\beta_{4}^{2} + 4k^{2})}
\label{expl4ng}
\end{equation}
An important conclusion which can be drawn from Eq.~(\ref{expl4ng})
is that the non-Gaussian character of the disorder produces an
enhancement of localisation. This increase may be negligible in the
present case, because the second-order Lyapunov exponent~(\ref{expl2})
never vanishes; that is not the case, however, when long-range
correlations of the potential come into play, as we will discuss in
the next section.

Considered together, Eqs.~(\ref{expl4g}) and~(\ref{expl4ng}) show that
the Gaussian and non-Gaussian parts of the fourth-order Lyapunov
exponent are both decreasing functions of the wavevector $k$, with the
same asymptotic behaviour for large values of $k$. This implies that the
relative importance of the two terms is determined only by the relative
magnitude of the cumulant strengths $\sigma_{2}$ and $\sigma_{4}$ and
of the decay constants $\beta_{2}$ and $\beta_{4}$; when the
corresponding parameters are of of the same order of magnitude, neither
component of the fourth-order Lyapunov exponent dominates the other.
Comparing Eq.~(\ref{expl2}) with~(\ref{expl4g}) and~(\ref{expl4ng}),
one can also observe that for large values of $k$ the second-order
Lyapunov exponent decays as $\lambda_{2}(k) \propto 1/k^{4}$, whereas
the fourth-order correction vanishes as $\lambda_{4} \propto 1/k^{8}$.
This shows how the Born approximation becomes more and more accurate
as the energy of the electrons is increased, in agreement with the
general principle that electrons of higher energy are less sensitive
to the details of the random potential.

Eqs.~(\ref{expl4g}) and~(\ref{expl4ng}) also show that the both parts
of the fourth-order Lyapunov exponent tend to vanish when the
correlation lengths $\beta_{2}^{-1}$ and $\beta_{4}^{-1}$ tend
simultaneously to infinity, as one can expect considering that the
localisation effects become less pronounced as the range of disorder
correlations is increased. Note, however, that the two localisation
lengths need not be equal; in particular, it is possible to consider
the case of a strongly correlated potential for which the correlation
function is constant in space (i.e., $\beta_{2} =0$) and the
spatial variation of disorder only shows up through the fourth
cumulant~(\ref{expcum}) with $\beta_{4} > 0$.
In this case the terms~(\ref{expl2}) and~(\ref{expl4g}) vanish and
the inverse localisation length coincides with the non-Gaussian
component~(\ref{expl4ng}).

Finally, we observe that the two parts~(\ref{expl4g}) and~(\ref{expl4ng})
of the fourth-order Lyapunov exponent tend to zero in the white-noise
limit, in agreement with the results of the previous subsection, as can
be seen by taking the limits $\beta_{2} \rightarrow \infty$ and
$\beta_{4} \rightarrow \infty$ (with the constraints
$\int \chi_{2}(x) dx = \tilde{\sigma}_{2}$ and
$\int \Delta_{4}(x_1,x_2,x_3) d^{3}x = \tilde{\sigma}_{4}$).

\section{Disorder with long-range correlations}
\label{lrc}

In this section, we focus our attention on the case of disorder with
long-range correlations; more precisely, we consider a potential with
correlation function of the form
\begin{equation}
\chi_{2}(x) = \sigma_{2} \frac{\sin(2k_{c}x)}{x} .
\label{lrchi}
\end{equation}
The correlator~(\ref{lrchi}) is substantially different from those
considered in the previous section, because it lacks any finite length
scale beyond which it becomes negligible; as we will see, this feature
strongly alters the localisation properties of the electronic
wavefunctions.
To enhance the physical understanding of the problem, we will analyse
first the physical properties which descend from the specific
form~(\ref{lrchi}) of the two-point correlation function,
postponing the discussion of non-Gaussian effects until
the end of this section.

As a consequence of the slow decay of the function~(\ref{lrchi}), great
attention must be paid when applying the formalism developed in
Sec.~\ref{pertexp} to the present case. In fact, in the derivation of
the general results of Sec.~\ref{pertexp} we avoided any mathematical
inconsistency by conveniently assuming the existence of a finite
correlation length; one could therefore question the chevalier
application of the general formulae~(\ref{lambda2}) and~(\ref{l4g}) to
a case in which no correlation length can be properly individuated and
condition~(\ref{lc}) is not satisfied.
To avoid these difficulties, we will study long-range correlators of
the form~(\ref{lrchi}) as the limit case for $\beta \rightarrow 0^{+}$
of the more general correlation function
\begin{equation}
\chi_{2}(x) = \sigma_{2} \frac{\sin(2k_{c}x)}{x}
\exp \left( -\beta x \right) .
\label{betachi}
\end{equation}
Since the correlator~(\ref{betachi}) decays exponentially for $\beta > 0$,
the results of Sec.~\ref{pertexp} can be safely used for this model; the
localisation properties for the particular case~(\ref{lrchi}) can then
be deduced by discussing the limit form of the Lyapunov exponent
for $\beta \rightarrow 0^{+}$.

Having defined our approach, we can proceed to derive the second-order
Lyapunov exponent for the correlation functions~(\ref{betachi})
and~(\ref{lrchi}). To simplify the mathematical form of the results
and get rid of absolute value signs, in the rest of this section we will
suppose that $k > 0$. This is not a restrictive hypothesis, because the
wavevector $k$ enters model~(\ref{mod}) only via the energy $E = k^{2}$
which is an even function of $k$; consequently, the Lyapunov exponent is
also even in $k$.

\begin{figure}[htp]
\begin{center}
\caption{Second-order Lyapunov exponent}
\label{l2fig}
\epsfig{file=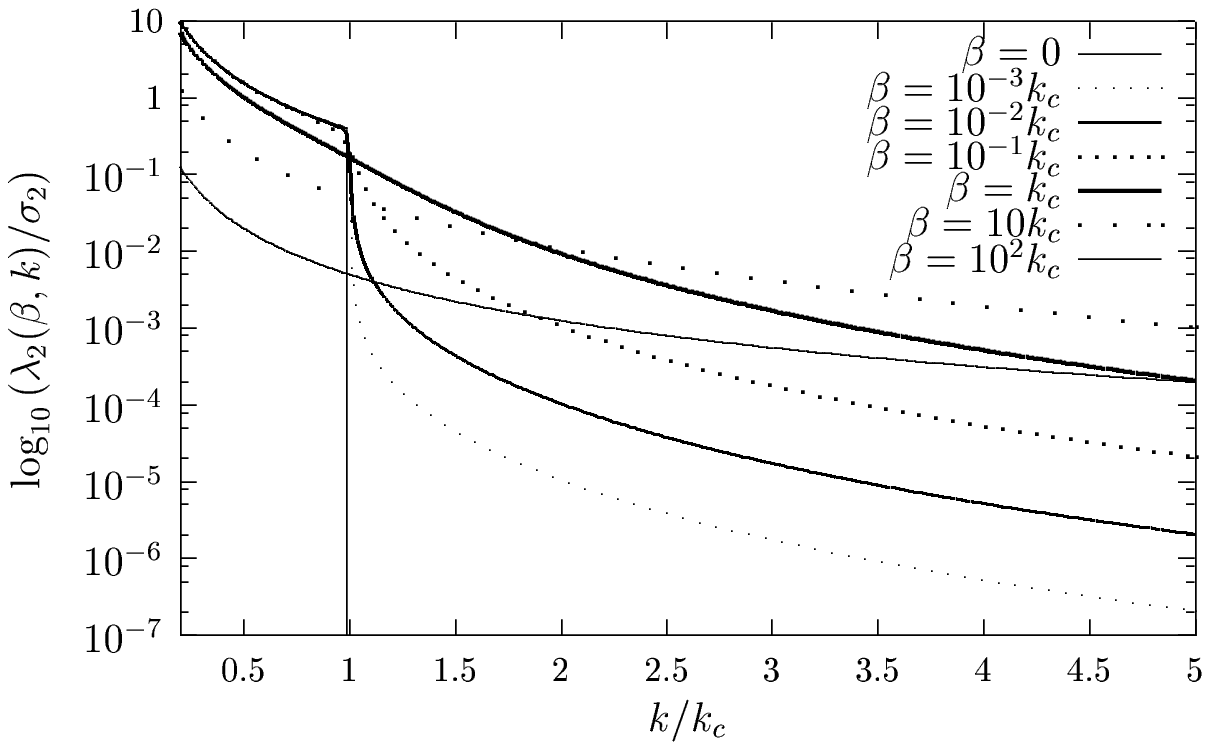,width=5in,height=3in}
\end{center}
\end{figure}

Inserting~(\ref{betachi}) in Eq.~(\ref{lambda2}), one obtains
\begin{equation}
\lambda_{2}(\beta, k) = \frac{\sigma_{2}}{8k^{2}} \left\{ \arctan \left[
\frac{4 \beta k_{c}}{\beta^{2} + 4 (k^{2} - k_{c}^{2})} \right]
+ \pi \theta \left( 4k_{c}^{2} - 4k^{2} -\beta^{2} \right)\right\}
\label{betal2}
\end{equation}
where $\theta(x)$ is the step function defined as
\begin{displaymath}
\theta(x) = \left\{ \begin{array}{ccl}
                     1  & \mbox{ if} & x > 0 \\
                     0  & \mbox{ if} & x \le 0 \\
               \end{array} \right. .
\end{displaymath}
In the limit $\beta \rightarrow 0^{+}$ expression~(\ref{betal2}) becomes
\begin{equation}
\lambda_{2}(k) = \left\{ \begin{array}{ccl}
    \displaystyle \frac{\sigma_{2} \pi}{8k^{2}} & \mbox{for} & 0<k<k_{c} \\
                                    0           & \mbox{for} &  k_{c}<k  \\
                          \end{array} \right.
\label{lrl2}
\end{equation}
The behaviour of the second-order Lyapunov exponent~(\ref{betal2})
and~(\ref{lrl2}) is represented in Fig.~\ref{l2fig}.

Note that the limit $\beta \rightarrow 0^{+}$ entails a sharp
qualitative change in both the correlation function and the
second-order Lyapunov exponent. For the correlation function, the
limit $\beta \rightarrow 0^{+}$ brings about a transition from an exponential
to a power-law decay; correspondingly, the strictly positive
Lyapunov exponent~(\ref{betal2}) is converted in the inverse localisation
length~(\ref{lrl2}) which vanishes when the wavevector $k$ exceeds the
critical value $k_{c}$.
Comparing expressions~(\ref{betal2}) and~(\ref{lrl2}), therefore, one
can conclude that the passage from a disorder with finite-range
correlations to a disorder with long-range correlations causes the
emergence of a continuum of delocalised states, separated by a
{\em mobility edge} from the localised ones~\cite{Izr99}.

This is obviously an important conclusion, because it shows that
a kind of Anderson transition can occur in 1D disordered models;
however, it is not completely satisfactory, because it is valid only
within the limits of the second-order approximation.
More precisely, one cannot label as ``extended'' the states for
$k > k_{c}$ without a certain margin of ambiguity, because the
fact that the second-order term~(\ref{lrl2}) vanishes does not guarantee
that the same be true for all higher-order terms of the Lyapunov
exponent. It is therefore impossible to ascertain on the basis of
Eq.~(\ref{lrl2}) whether the electronic states for $k > k_{c}$ are
really extended or, rather, localised in a different form or on a
much larger spatial scale than the states for $k < k_{c}$.
To shed light on this point, it is useful to determine the fourth-order
Lyapunov exponent~(\ref{l4g}) in the case of long-range correlations
of the form~(\ref{lrchi}). Once more, we will first compute the
fourth-order term~(\ref{l4g}) for the correlator~(\ref{betachi}) and
then we will use this result to study the limit $\beta
\rightarrow 0^{+}$. 

When the two-point correlation function has the form~(\ref{betachi}), 
the functions~(\ref{phifun}) become
\begin{eqnarray*}
\varphi_{c}(k,x) & = & \frac{\sigma_{2}}{2}
\mbox{Im} \left\{ E_{1} \left[ \beta x - i2(k+k_{c})x \right] -
E_{1} \left[ \beta x -i2(k-k_{c})x \right] \right\} \\
\varphi_{s}(k,x) & = & \frac{\sigma_{2}}{2}
\mbox{Re} \left\{ -E_{1} \left[ \beta x - i2(k+k_{c})x \right] +
E_{1} \left[ \beta x -i2(k-k_{c})x \right] \right\}, \\
\end{eqnarray*}
where $E_{1}(z)$ is the exponential integral defined in the complex plane
by
\begin{eqnarray*}
E_{1} (z) = \int_{z}^{\infty} \frac{e^{-t}}{t} dt & \mbox{with} &
|\arg z| < \pi \\
\end{eqnarray*}
(see, e.g.,~\cite{Abr65}).
Correspondingly, the Gaussian part of the fourth-order Lyapunov
exponent takes the form
\begin{equation}
\begin{array}{ccl}
\lambda_{4}^{(G)}(\beta,k) & = & 
\displaystyle
\frac{\sigma_{2}^{2}}{4k^{4}} \left\{ \frac{1}{16}
\left[ \frac{\beta}{\beta^{2} + 4(k-k_{c})^{2}} - \frac{\beta}{\beta^{2}
+ 4(k+k_{c})^{2}} \right. \right. \\
& + & \displaystyle
\frac{1}{k} \left. \left( \alpha_{1}(k) + \frac{\pi}{2}
s_{3}(k) \right) \right] \ln \frac{\beta^{2} + 4(k+k_{c})^{2}}
{\beta^{2} + 4(k-k_{c})^{2}} \\
& - & \displaystyle
\frac{\beta}{8} \left[ \frac{1}{\beta^{2} + 4k_{c}^{2}} +
\frac{1}{\beta^{2} + 4(k+k_{c})^{2}} \right]
\ln \frac{\beta^{2}+(k+2k_{c})^{2}}{\beta^{2}+k^{2}} \\
& - & \displaystyle
\frac{\beta}{8} \left[ \frac{1}{\beta^{2} + 4k_{c}^{2}} +
\frac{1}{\beta^{2} + 4(k-k_{c})^{2}} \right]
\ln \frac{\beta^{2}+(k-2k_{c})^{2}}{\beta^{2}+k^{2}} \\
& - & \displaystyle
\frac{1}{2} \frac{k+k_{c}}{\beta^{2} + 4(k+k_{c})^{2}} \left[ \alpha_{1}(k)
+ \alpha_{3}(k) + \frac{\pi}{2} s_{3}(k) \right] \\
& + & \displaystyle
\frac{1}{2} \frac{k-k_{c}}{\beta^{2} + 4(k-k_{c})^{2}} \left[ \alpha_{1}(k)
+ \alpha_{4}(k) + \frac{\pi}{2} s_{3}(k) + \pi s_{2}(k) \right] \\
& + & \displaystyle
\left. \frac{k_{c}}{\beta^{2} + 4k_{c}^{2}} \left[ 2 \alpha_{1}(k) -
\alpha_{2}(k) + \pi s_{3}(k) - \frac{\pi}{2} s_{1}(k) \right]
+ F_{\beta}(k) \right\} \\
\end{array}
\label{betal4}
\end{equation}
where the functions $\alpha_{i}(k)$ are defined as
\begin{eqnarray*}
\alpha_{1}(k) & = & \frac{1}{2} \arctan \frac{4 \beta k_{c}}{\beta^{2} +
4(k^{2}-k_{c}^{2})} \\
\alpha_{2}(k) & = & \frac{1}{2} \arctan \frac{4 \beta k_{c}}{\beta^{2} +
k^{2} - 4k_{c}^{2}} \\
\alpha_{3}(k) & = & \arctan \frac{2 \beta k_{c}}{\beta^{2} + k^{2}
+ 2kk_{c}} \\
\alpha_{4}(k) & = & \arctan \frac{2 \beta k_{c}}{\beta^{2} + k^{2}
- 2kk_{c}} \\
\end{eqnarray*}
and the symbols $s_{i}(k)$ represent the step-functions
\begin{eqnarray*}
s_{1}(k) & = & \left\{ \begin{array}{ccl}
1  & \mbox{ if} & \beta^{2} + k^{2} - 4k_{c}^{2} < 0 \\
0  &             & \mbox{otherwise} \\
               \end{array} \right. \\
s_{2}(k) & = & \left\{ \begin{array}{ccl}
1  & \mbox{ if} & \beta^{2} + k^{2} - 2kk_{c} < 0 \\
0  &             & \mbox{otherwise} \\
               \end{array} \right. \\
s_{3}(k) & = & \left\{ \begin{array}{ccl}
1  & \mbox{ if} & \beta^{2} + 4(k^{2} - k_{c}^{2}) < 0 \\
0  &             & \mbox{otherwise} \\
               \end{array} \right. . \\
\end{eqnarray*}
Finally, the function $F_{\beta}(k)$ which appears in the r.h.s. of
Eq.~(\ref{betal4}) is defined by the integral representation
\begin{equation}
F_{\beta}(k) = \frac{1}{\sigma_{2}^{2}k} \int_{0}^{\infty} \chi_{2}(x)
\left[ \varphi_{c}(0,0) - \varphi_{c}(0,x) \right] \sin (2kx) dx .
\label{fk}
\end{equation}
For $k > 2k_{c}$ the function~(\ref{fk}) can be expressed through
the series
\begin{equation}
\begin{array}{ccl}
F_{\beta}(k) & = & \displaystyle
\frac{1}{2k} \sum_{n=1}^{\infty} \frac{(-1)^{n}}{n^{2}} \left\{
\left[ \frac{\beta^{2} + 4k_{c}^{2}}{\beta^{2} + 4(k+k_{c})^{2}}
\right]^{\frac{n}{2}} \cos(n \delta_{1}) \right.\\
& - & \displaystyle \left.
\left[ \frac{\beta^{2} + 4k_{c}^{2}}{\beta^{2} + 4(k-k_{c})^{2}}
\right]^{\frac{n}{2}} \cos(n \delta_{2}) \right\}
\sin(n \delta_{0})
\end{array}
\label{serie1}
\end{equation}
where
\begin{eqnarray*}
\delta_{0} & = & \arctan \frac{2k_{c}}{\beta} \\
\delta_{1} & = & \arctan \frac{2(k+k_{c})}{\beta} \\
\delta_{2} & = & \arctan \frac{2(k-k_{c})}{\beta} .\\
\end{eqnarray*}
For $k < \sqrt{k_{c}^{2} + \beta^{2}} - k_{c}$, the function F(k) can
be represented in the alternative form
\begin{displaymath}
F_{\beta}(k) = \frac{1}{4k} \arctan \left( \frac{2k_c}{\beta} \right)
\ln \frac{\beta^{2} + 4(k+k_c)^{2}}{\beta^{2} + 4(k-k_c)^{2}}
+ G_{\beta}(k)
\end{displaymath}
where $G_{\beta}(k)$ is the series
\begin{equation}
G_{\beta}(k) = \frac{1}{2k} \sum_{n=1}^{\infty} \frac{(-1)^{n}}{2n}
c_{n} \left\{ \left[ \frac{(k+k_c)^{2}}{\beta^{2} +k_c^{2}} \right]^{n}
- \left[ \frac{(k-k_c)^{2}}{\beta^{2} +k_c^{2}} \right]^{n}\right\}
\label{serie2}
\end{equation}
with coefficients
\begin{displaymath}
c_{n} = \sum_{l=0}^{\infty} \frac{1}{2n+l}
\left( \frac{\beta}{2\sqrt{\beta^{2}+k_{c}^{2}}} \right)^{l}
\sin \left[ (2n+l) \arctan \left( \frac{k_c}{\beta} \right) \right] .
\end{displaymath}
Note that, when $\beta \ge \sqrt{8}k_{c}$, the convergence regions of
the series~(\ref{serie1}) and~(\ref{serie2}) overlap, so that there is
no need to use the integral representation~(\ref{fk}) to compute
the values of the function $F_{\beta}(k)$; the integral~(\ref{fk}) must
be evaluated only if $\beta < \sqrt{8}k_{c}$ and for those values of $k$
comprised in the interval
$\sqrt{k_{c}^{2} + \beta^{2}} - k_{c} < k < 2k_{c}$.

\begin{figure}[htp]
\begin{center}
\caption{Fourth-order Lyapunov exponent for $k<2k_{c}$ and $\beta \ge k_{c}$}
\label{l4fig1}
\epsfig{file=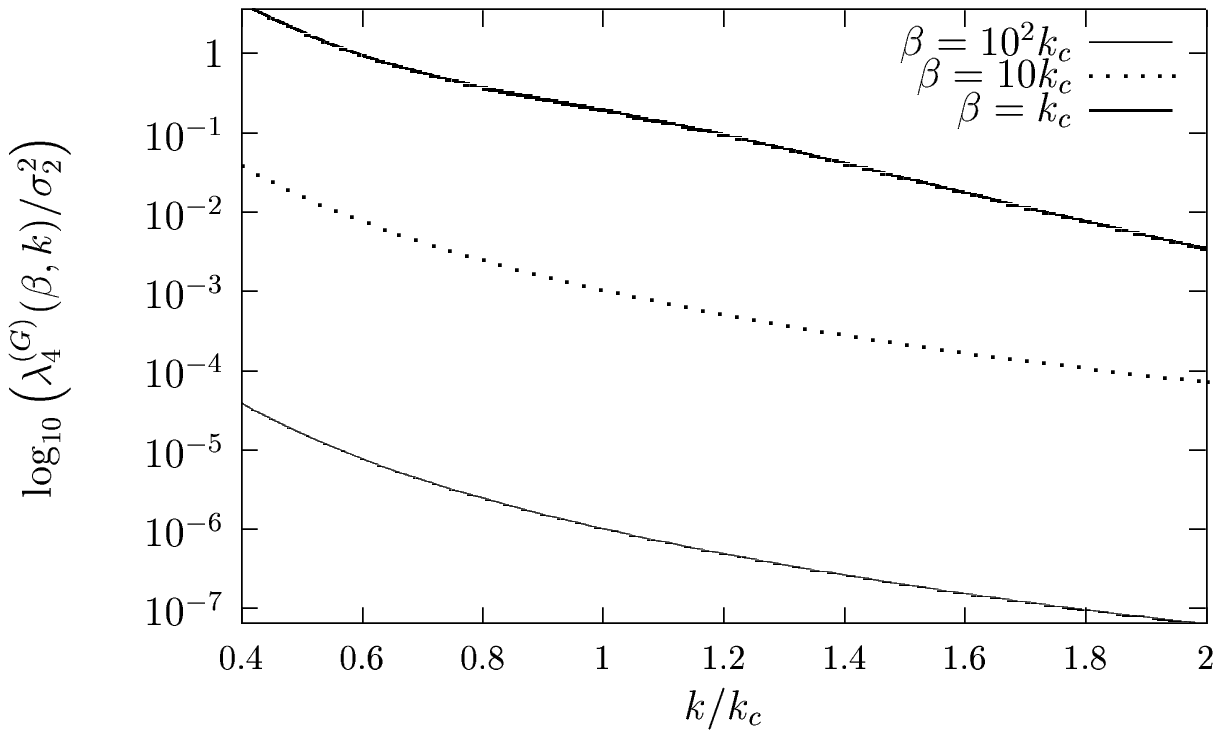,width=5in,height=3in}
\end{center}
\end{figure}

\begin{figure}[htp]
\begin{center}
\caption{Fourth-order Lyapunov exponent for $k<2k_{c}$ and $\beta \le k_{c}$}
\label{l4fig2}
\epsfig{file=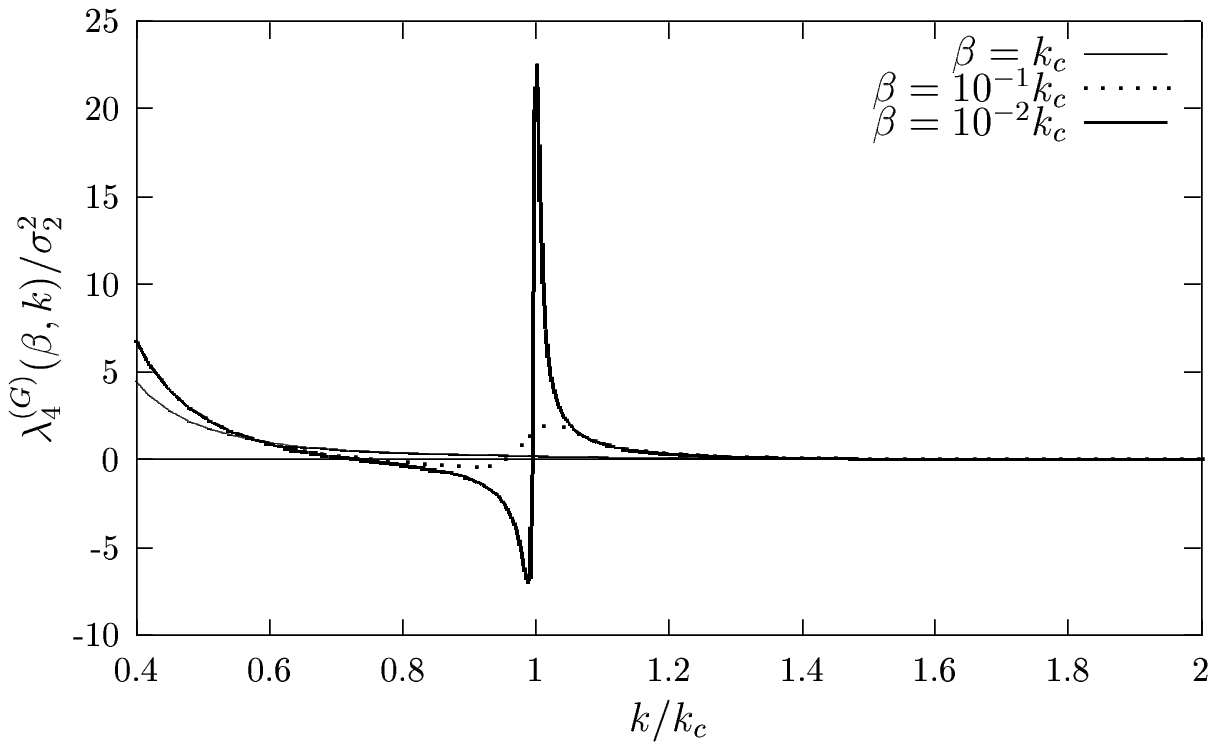,width=5in,height=3in}
\end{center}
\end{figure}

\begin{figure}[htp]
\begin{center}
\caption{Fourth-order Lyapunov exponent for $k>2k_{c}$}
\label{l4fig3}
\epsfig{file=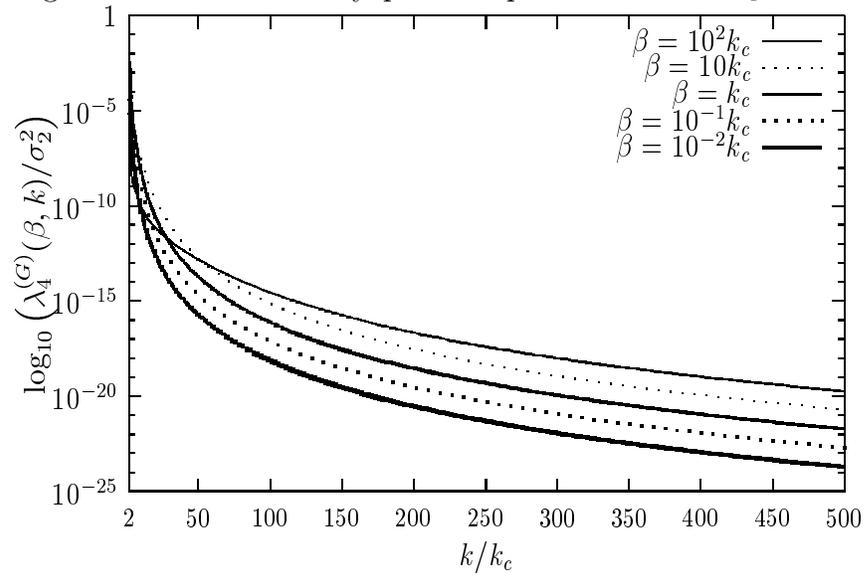,width=5in,height=3in}
\end{center}
\end{figure}

The behaviour of the fourth-order Lyapunov exponent~(\ref{betal4}) is
represented for $k < 2k_{c}$ in Figs.~\ref{l4fig1} and~\ref{l4fig2},
and for $k>2k_{c}$ in Fig.~\ref{l4fig3}.
As shown by the figures, for large values of $\beta$ the fourth-order
Lyapunov exponent is a smooth and decreasing function of $k$, assuming
small but positive values everywhere (except, of course, than in the
low-energy limit $k \rightarrow 0$, where the results are not valid
because the weak disorder condition~(\ref{weakdis}) is no longer
satisfied). Physically this means that, when the correlation range of
the disordered potential is short, the fourth-order Lyapunov
exponent~(\ref{betal4}) is a negligible correction to the second-order
term~(\ref{betal2}).
The behaviour of $\lambda_{4}^{(G)}(\beta,k)$ becomes more complex
for small values of $\beta$. First, the function looses its monotonic
behaviour and even starts to assume negative values in an interval
comprised within the range $0 < k < k_{c}$.
Note that the fourth-order term can become negative as long as the
second-order term is positive; physically, this means that the
fourth-order term reduces the localisation effect.
Finally, for $\beta \ll k_{c}$, the function $\lambda_{4}^{(G)}(\beta,k)$
develops a pronounced negative minimum in a left neighbourhood of $k_{c}$
and a sharp positive peak in a right neighbourhood $k_{c}$.
For values of $k$ close to $k_{c}$, therefore, the fourth-order term
can represent a conspicuous correction to the leading term when
$\beta \ll k_{c}$; since in this limit the second-order term tends
to assume the discontinuous form~(\ref{lrl2}), one can conclude that
the effect of the fourth-order correction is that of smoothing
the Lyapunov exponent in the region $k \simeq k_{c}$.

\begin{figure}[htp]
\begin{center}
\caption{Fourth-order Lyapunov exponent for $\beta=0$}
\label{l4fig4}
\epsfig{file=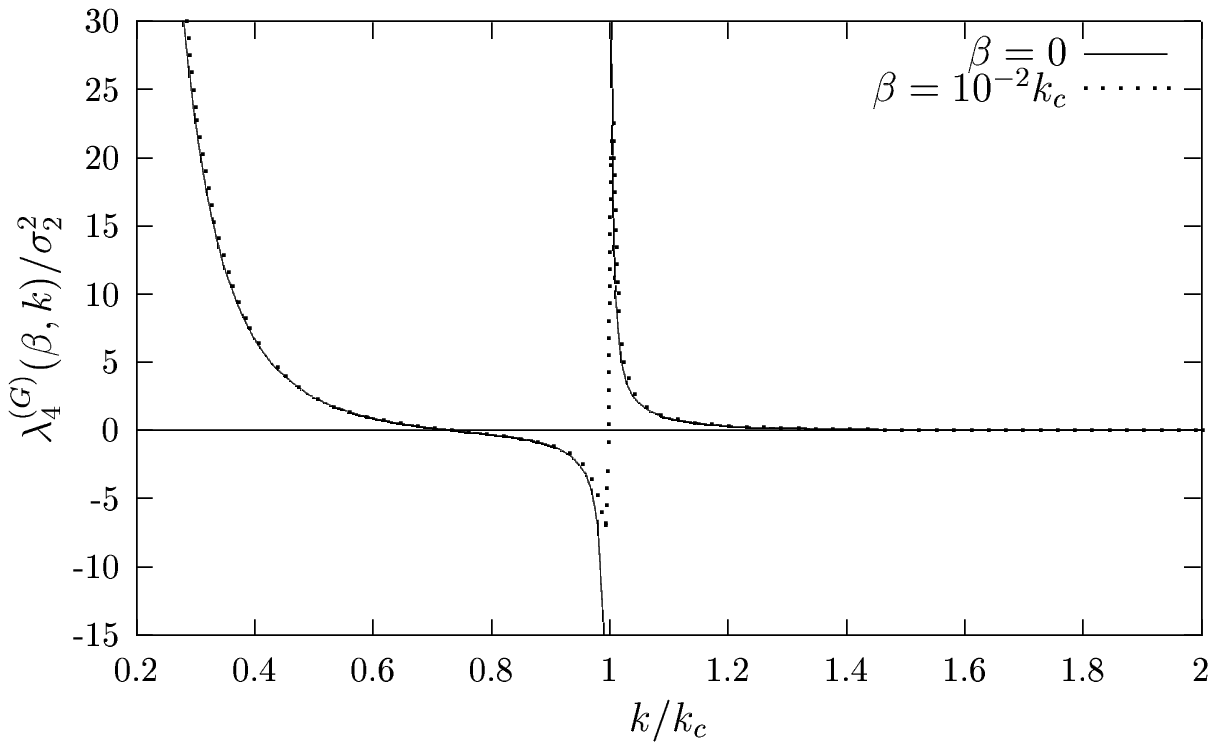,width=5in,height=3in}
\end{center}
\end{figure}

\begin{figure}[htp]
\begin{center}
\caption{Fourth-order Lyapunov exponent for $\beta=0$ and $k>k_{c}$}
\label{l4fig5}
\epsfig{file=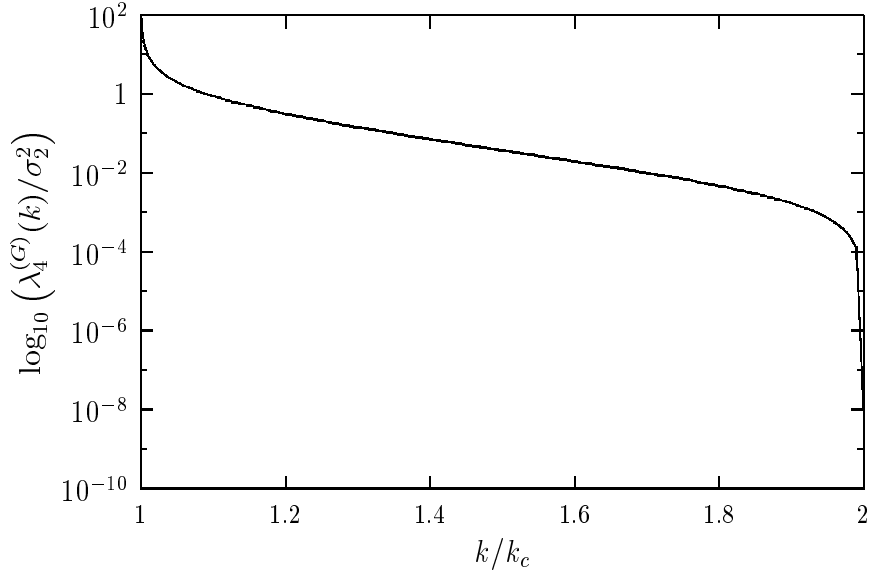,width=5in,height=3in}
\end{center}
\end{figure}

The expression of the fourth-order Lyapunov exponent~(\ref{betal4}) is
quite complicated, but it simplifies significantly in the case of
greatest interest, i.e., in the limit $\beta \rightarrow 0^{+}$, when
it takes the form
\begin{equation}
\lambda_{4}^{(G)}(k) = \left\{ 
\begin{array}{lcl}
\begin{array}{l}
\displaystyle
\frac{\sigma_{2}^{2}}{4k^{4}} \left[ \frac{\pi}{32k}
\ln \left( \frac{k+k_{c}}{k-k_{c}} \right)^{2} \right.\\
\displaystyle \left.
- \frac{\pi}{8} \frac{k^{2}+kk_{c}+k_{c}^{2}}{k_{c}(k_{c}^{2}
-k^{2})} + F_{0}(k) \right] 
\end{array}
& \mbox{for} & 0 < k < k_{c} \\
\displaystyle
\frac{\sigma_{2}^{2}}{4k^{4}}
\left[ \frac{\pi}{8} \frac{2k_{c}-k}{k_{c}(k-k_{c})}
+ F_{0}(k) \right] & \mbox{for} & k_{c} < k < 2k_{c} \\
0  & \mbox{for} & 2 k_{c} < k \\
\end{array} \right. .
\label{lrl4}
\end{equation}
The function~(\ref{lrl4}) is plotted in Figs.~\ref{l4fig4}
and~\ref{l4fig5}.
In Fig.~\ref{l4fig4} the graph of $\lambda_{4}^{(G)}(\beta,k)$ with
$\beta=10^{-2}k_{c}$ is also plotted for comparison. As can be seen
from Fig.~\ref{l4fig4}, the fourth-order Lyapunov exponent~(\ref{lrl4})
displays a qualitative behaviour close to that of the corresponding
term~(\ref{betal4}) for small values of $\beta$. There are two main
differences, though: first, the fourth-order Laypunov exponent~(\ref{lrl4})
is strictly zero for $k>2k_{c}$ and, second, it diverges for $k \rightarrow
k_{c}$.

This divergence must be disregarded, however, because the limit $\beta
\rightarrow 0^{+}$ cannot be legitimately taken in a neighbourhood of
$k = k_{c}$. To understand this point we recall that, as pointed out
in Sec.~\ref{pertexp}, the existence of the asymptotic
generators~(\ref{asymgen}) cannot be taken for granted when one drops
the assumption that a finite correlation length exists. In the present
case, a problem of this kind arises for the second-order asymptotic
generator~(\ref{k2}) which cannot be defined for $k=k_{c}$ in the
limit $\beta \rightarrow 0^{+}$.
In fact, as can be deduced from formula~(\ref{k231}), the matrix element
$({\bf{K}}_{2})_{31}$ of the second-order asymptotic generator is
proportional to the sine transform of the correlation function.
The sine transform of the correlator~(\ref{betachi}) is a well-behaved
function as long as $\beta > 0$; in the limit $\beta \rightarrow 0^{+}$,
however, it diverges for $k=k_{c}$.
As a consequence, the general formulae~(\ref{l4g}) and~(\ref{l4ng}) can
be applied to the study of disorder with long-range
correlations~(\ref{lrchi}) only for wavevectors $k$ which are not close
to the critical value $k_{c}$.
We can thus conclude that formula~(\ref{lrl4}) describes the correct
behaviour of the fourth-order Lyapunov exponent everywhere, except that
in a small neighbourhood of $k_{c}$.
(Strictly speaking, the previous analysis also invalidates the
second-order result~(\ref{lrl2}) for $k \simeq k_{c}$; this expression,
however, descends from formula~(\ref{lambda2}), which has a
broader range of validity since it can be derived with alternative
methods~\cite{Lif88} that do not require the existence of the asymptotic
generator ${\bf{K}}_{2}$.)

Having established the range of validity of the expression~(\ref{lrl4}),
we can now comment the physical meaning of the result. The first
remarkable aspect of the fourth-order Lyapunov exponent~(\ref{lrl4})
is that it takes positive values in the interval $k_{c} < k < 2k_{c}$
where the second-order term~(\ref{lrl2}) vanishes. This means that
the inverse localisation length, which is a quadratic function of the
potential for $k < k_{c}$, assumes a quartic form when the
wavevector exceeds the critical value $k_{c}$.
This feature clarifies the nature of the transition which occurs
at $k = k_{c}$: when the threshold is crossed, the electronic states
continue to be exponentially localised, but over spatial scales which
are much larger than those typical of the states with $k < k_{c}$.
The transition at $k_{c}$, therefore, does not bring a change from
exponential localisation to complete delocalisation, but an increase of
order $O(1/\varepsilon^{2})$ in the spatial extension of the electronic
states. Since we are considering the case of weak disorder, $\varepsilon
\ll 1$, this increase can be huge: the bold use of terms like ``mobility
edge'' and ``delocalisation transition'' adopted in Ref.~\cite{Izr99} is
thus fully justified.

A second relevant feature of the fourth-order Lyapunov
exponent~(\ref{lrl4}) is that it vanishes for $k > 2k_{c}$. This
implies that, when the potential displays long-range correlations
of the form~(\ref{lrchi}), a second transition takes place at
$k = 2k_{c}$, with the electrons becoming even more weakly localised.
The spatial extension of the electronic states beyond this second
threshold cannot be estimated within the fourth-order approximation, which
allows the only conclusion that for $k>2k_{c}$ the inverse localisation
length must be an infinitesimal of order $O(\varepsilon^{6})$ or higher.

The emergence of a second threshold at $k = k_{c}$ is an effect of the
Gaussian nature of the potential that we have been considering up to
now. The situation is different for non-Gaussian potentials, because
in this case localisation is enhanced by the cumulant-generated
term~(\ref{l4ng}). The specific form of the non-Gaussian term~(\ref{l4ng})
depends on that of the fourth cumulant of the disorder; to discuss
a concrete example, we will consider the case of a random potential
with correlation function and fourth cumulant of the forms~(\ref{lrchi})
and~(\ref{expcum}), respectively.
The fact that the correlation function has the form~(\ref{lrchi}) implies
that all moments of the potential decay slowly, so that it is correct
to speak of disorder with long-range correlations at all orders.
On the other hand, the non-zero fourth cumulant~(\ref{expcum})
makes the fourth moment of the potential different from what it would
be in the Gaussian case. Since the cumulant~(\ref{expcum}) decays
exponentially, the fourth moment of the potential significantly differs
from its Gaussian form only in a limited spatial range (which can actually
be very small if $\beta_{4}$ is large); yet this difference, no matter
how small, is enough to induce the localisation of all electronic
eigenstates with $k > 2 k_{c}$ with inverse localisation length equal
to~(\ref{expl4ng}).
We can therefore conclude that, in the case of disorder with long-range
correlations of the form~(\ref{lrchi}), non-Gaussian potentials generate
a stronger localisation of the electrons that their Gaussian counterparts.

To conclude this section, we remark all the results obtained so far
for the localisation of electrons in a disordered solid can be transposed
to any other model described by Eq.~(\ref{mod}): in particular, the
results of this section can be applied to the dynamics of the stochastic
oscillator~(\ref{randosc}). In the language appropriate to this model,
the results of this section can be expressed by saying that, when the
stochastic perturbation of the frequency displays long-range
temporal correlations of the form~(\ref{lrchi}), the behaviour of the
oscillator can be dramatically altered by letting the unperturbed
frequency $k$ cross some critical thresholds.
Specifically, the oscillator is energetically unstable, with energy
growing with the rate~(\ref{lrl2}), for frequencies lower than $k_{c}$.
As soon as the unperturbed frequency exceeds the value $k_{c}$, however,
the energy growth rate drops by a factor $O(\varepsilon^{2})$ so that the
energetic instability of the oscillator manifests itself only on long time
scales. Finally, for $k > 2k_{c}$ the oscillator becomes even more stable,
with a further reduction of the rate of energy growth which may be at least
of order $O(\varepsilon^{2})$ or of a simple factor $O(\varepsilon^{0})$
depending on whether the noise is Gaussian or not.

\section{Conclusions}
\label{conclu}

In this paper we studied the localisation properties of 1D models
with weak disorder, focusing our attention on the role played
by spatial correlations of the random potential in shaping the
structure of the electronic states.
Using a perturbative approach based on a cumulant expansion technique,
we were able to derive both the standard second-order expression for the
inverse localisation length and a new result for the fourth-order
correction.
The knowledge of the fourth-order term of the Lyapunov exponent makes
possible to investigate the delocalisation transition which takes place
when the disorder exhibits specific long-range correlations.
The analysis of Sec.~\ref{lrc} reveals that this transition consists
in a sharp change of the scaling law for the inverse localisation
length: when the electronic wavevector reaches a critical value, the
Lyapunov exponent switches from a quadratic to a quartic dependence on
the disorder strength $\varepsilon$ and, correspondingly, the spatial
extension of the electronic states increases by a factor
$O(1/\varepsilon^{2})$.
In the case of Gaussian potentials, a second critical value of the
wavevector exists, beyond which the electronic states become even
more delocalised, with Lyapunov exponent $\lambda = o(\varepsilon^{4})$
for $\varepsilon \rightarrow 0$.

This additional transition is absent in the case of non-Gaussian
potentials, which cannot be distinguished from their Gaussian counterparts
in the second-order approximation but whose specific features emerge
when the description is brought to the refined fourth-order level.
Specifically, non-Gaussian potentials show their effect through
the addition to the Gaussian Lyapunov exponent of an extra term
proportional to the fourth cumulant of the random potential.
This cumulant-generated term strengthens the localisation effect and
maintains the quartic scaling of the Lyapunov exponent.
Loosely speaking, the requirement that the disorder be Gaussian seems
to act as a constraint which reduces the degree of disorder and
therefore favours the delocalisation of the electronic states,
in agreement with the general principle which links the degree
of randomness with that of localisation of the electronic states.

From this point of view, this paper can be seen as a further step
towards the construction of a hierarchy of disorders according to
the strength of the localisation effects they produce in 1D models.
In totally disordered systems, for which the potential has no spatial
correlation, all electronic states are exponentially localised; when
the degree, if not the strength, of the disorder is somewhat reduced by
the presence of short-range correlations of the potential, a discrete set
of extended states may appear; finally, when the potential exhibits
specific long-range spatial correlations and thus the randomness of
the model is further diminished, a continuum of states with large
spatial extension emerges.
This paper clarifies the features of the extended states which
appear in the last case and shows that, in the class of disorders with
long-range correlations, Gaussian potentials seem to be ``less random''
than non-Gaussian ones.

\section{Acknowledgments}

The author wishes to thank S. Ruffo and A. Politi for useful discussions
and is grateful to F. M. Izrailev for valuable comments and suggestions.

\end{document}